# Assessing Fiscal Policy Effectiveness on Household Savings in Hungary, Slovenia, and the Czech Republic during the COVID-19 Crisis: A Markov Switching VAR Approach

**Author:**

**Tuhin G M Al Mamun**

Hannam University, Department of Economics,South Korea

20224130@gm.hannam.ac.kr

**ORCID**: 0009-0005-0275-6922

**Abstract**

The COVID-19 pandemic significantly disrupted household consumption, savings, and income across Europe, particularly affecting countries like Hungary, Slovenia, and the Czech Republic. This study investigates the effectiveness of fiscal policies in mitigating these impacts, focusing on government interventions such as spending, subsidies, revenue, and debt. Utilizing a Markov Switching Vector Auto regression (MS-VAR) model, the study examines data from 2000 to 2023, considering three economic regimes: the initial shock, the peak crisis, and the recovery phase. The results indicate that the COVID-19 shock led to a sharp decline in household consumption and income in all three countries, with Slovenia facing the most severe immediate impact. Hungary, however, showed the strongest recovery, driven by effective fiscal measures such as subsidies and increased government spending, which significantly boosted both household consumption and income. The Czech Republic demonstrated a more gradual recovery, with improvements observed in future-oriented consumption (IMPC). In conclusion, the study underscores the critical role of targeted fiscal interventions in mitigating the adverse effects of crises. The findings suggest that governments should prioritize timely and targeted fiscal policies to support household financial stability during economic downturns and ensure long-term recovery.

## 1. Introduction

The COVID-19 pandemic caused a severe economic crisis, disrupting financial stability and household income. Governments worldwide introduced fiscal measures to counteract the economic downturn. Policies such as subsidies, direct transfers, increased public spending, and tax relief were used to support households and maintain economic activity. However, the success of these measures varied across countries, depending on fiscal space, economic structure, and policy execution.

Central European nations, including the Czech Republic, Hungary, and Slovenia, adopted different fiscal strategies to stabilize household consumption and income. Some countries focused on direct financial support, while others emphasized expenditure restructuring and debt management. The effectiveness of these policies depended on how households responded to government interventions. Studying these responses helps policymakers design better fiscal strategies for future economic crises.

Households are central to economic stability. During crises, financial uncertainty forces them to adjust spending and savings behavior. If people reduce consumption and increase savings, businesses suffer, and economic recovery slows. Fiscal policies aim to encourage household spending while ensuring income stability. However, if fiscal policies fail to stimulate consumption, they may only increase public debt without generating economic growth.

In Czech Republic, Hungary, and Slovenia, governments implemented various fiscal interventions. Some countries saw strong consumer spending responses, while others faced higher precautionary savings, weakening policy effectiveness. Understanding the short-term and long-term effects of fiscal policies on household consumption is crucial for developing better crisis-response strategies.

Most research on fiscal policy effectiveness focuses on large economies like the United States and Western Europe. Less is known about how smaller Central European economies handled fiscal interventions during the COVID-19 pandemic. There is limited analysis of how government debt, spending, revenue, and subsidies influenced household financial behavior in these countries.

Another gap in existing research is the changing impact of fiscal policy across different economic conditions. Many studies assume that government interventions have a consistent effect over time, but household behavior shifts depending on economic uncertainty and recovery stages. A Markov Switching VAR (MS-VAR) model is well-suited for capturing these shifts, as it identifies different economic regimes and tracks how fiscal effectiveness evolves in response to crisis conditions.

This study fills these gaps by applying an MS-VAR model to analyze fiscal effectiveness during three distinct economic phases:

Regime 1: Initial Shock (2020) – Households responded to sudden economic restrictions and policy uncertainty.

Regime 2: Peak Crisis (2021) – Governments intensified fiscal interventions, but the effectiveness of these measures varied.

Regime 3: Recovery Phase (2022) – The economy began stabilizing, and fiscal policy adjustments influenced household consumption.

This study provides insights for governments, policymakers, and researchers by analyzing how different fiscal measures influenced household consumption and disposable income. The findings will help policymakers improve fiscal crisis management strategies by identifying the most effective interventions for supporting households during economic uncertainty.

This study aims to answer the following key questions:

1. What is the impact of COVID-19 on household consumption and disposable income across three regimes (initial, peak, and recovery)?
2. What is the effectiveness of government subsidies and transfers across the three countries?
3. What is the effectiveness of fiscal sustainability and household consumption across the three countries?
4. What is the effectiveness of government revenue and its impact in the three countries?
5. What is the effectiveness of government expenses and their impact in the three countries?

By answering these questions, this study provides a detailed assessment of fiscal policy effectiveness in Czech Republic, Hungary, and Slovenia during the COVID-19 crisis.

## 2. Literature Review

Fiscal policy plays a critical role in stabilizing economies during crises. Governments use public spending, tax reductions, and direct financial support to protect household consumption and disposable income. Studies have shown that fiscal interventions can mitigate income shocks and sustain consumer spending during economic downturns (Auerbach & Gorodnichenko, 2012). However, the effectiveness of fiscal policies depends on the structure of the economy, household financial behavior, and the extent of government intervention.

During the 2008 Global Financial Crisis, government stimulus programs helped maintain consumption levels in many developed economies (Blanchard et al., 2010). Direct transfers and subsidies had a stronger impact on spending compared to tax reductions, as lower-income households tend to have a higher marginal propensity to consume (MPC) (Parker et al., 2013). Similar trends were observed during the COVID-19 pandemic, where direct financial support helped sustain household expenditure patterns (Chetty et al., 2020). However, research on fiscal interventions in smaller Central European economies remains limited.

Households adjust their financial decisions based on economic uncertainty. Traditional economic theories, such as the Permanent Income Hypothesis (Friedman, 1957) and the Life-Cycle Hypothesis (Modigliani & Brumberg, 1954), suggest that consumers aim to smooth their consumption over time. However, in crises, many households shift toward precautionary savings, reducing overall demand (Carroll & Kimball, 2006).

Empirical studies show that households in financially stable economies respond more positively to fiscal stimulus compared to those in countries with high public debt (Jappelli & Pistaferri, 2014). In some cases, excessive fiscal spending can increase debt burdens, leading to long-term economic instability. Countries with limited fiscal space may struggle to sustain stimulus efforts without causing inflationary pressures (Corsetti et al., 2012).

In Central Europe, household behavior during COVID-19 was influenced by income stability, government transfers, and labor market conditions. Countries with strong wage subsidies and financial relief programs saw higher household spending retention, while those with weaker fiscal measures experienced increased savings rates and reduced consumption.

A key measure of fiscal policy effectiveness is the marginal propensity to consume (MPC), which reflects how much of an additional unit of income households spend rather than save. Studies indicate that MPC values vary by income group and economic conditions (Jappelli & Pistaferri, 2014). Low-income households tend to have higher MPCs, meaning they are more likely to spend government transfers immediately. In contrast, wealthier households save a larger share of fiscal benefits, reducing the short-term impact of stimulus measures (Kaplan & Violante, 2014).

The intertemporal marginal propensity to consume (IMPC) considers how households adjust their spending based on expected future income. If economic uncertainty is high, IMPC tends to be lower, as households save more in anticipation of future risks (Attanasio & Weber, 2010). Understanding these behavioral responses is essential for assessing the success of fiscal interventions in the Czech Republic, Hungary, and Slovenia.

One of the biggest concerns of expansionary fiscal policy is long-term sustainability. Governments must balance short-term economic support with long-term financial stability. Excessive public debt can lead to higher borrowing costs, inflationary pressures, and reduced policy flexibility (Reinhart & Rogoff, 2010).

Some studies suggest that fiscal consolidation strategies—such as reducing deficits after a crisis—can help stabilize economies in the long run (Alesina & Ardagna, 2010). However, premature fiscal tightening can slow recovery and increase unemployment. The challenge is finding the right balance between stimulus and sustainability.

In Central Europe, countries with stronger fiscal positions before the crisis were able to provide larger stimulus packages without facing immediate debt risks. Others, particularly those with high pre-existing debt, had limited fiscal space, restricting their ability to support households effectively.

The fiscal responses of Czech Republic, Hungary, and Slovenia to the COVID-19 crisis varied significantly:

- Czech Republic: Implemented strong fiscal stimulus, including direct income support, subsidies, and tax deferrals. The government maintained a balance between stimulus and fiscal discipline, ensuring moderate debt increases while sustaining household consumption.
- Hungary: Relied on a mix of wage subsidies, loan repayment moratoriums, and public investment projects. However, rising government debt and inflation concerns limited the effectiveness of fiscal policies.
- Slovenia: Adopted moderate fiscal interventions, focusing on employment protection programs and targeted subsidies. The policy was effective in stabilizing income, but limited spending incentives led to higher precautionary savings among households.

The effectiveness of these fiscal policies remains debated, making an empirical analysis necessary to determine their impact on household financial behavior.

There is limited research on how fiscal policies influenced household savings and consumption in Central European economies during the COVID-19 crisis. Existing studies mostly focus on large Western economies. Few empirical analyses examine how fiscal interventions performed under changing economic conditions across different regimes.

This study contributes to economic research by:

1. Applying a Markov Switching VAR (MS-VAR) model to capture regime-specific fiscal policy effectiveness in Central Europe.
2. Providing comparative evidence on fiscal policy impacts in Czech Republic, Hungary, and Slovenia.

3. Assessing fiscal sustainability by evaluating how public debt and revenue policies affected household financial behavior.

By addressing these gaps, this study offers valuable insights for policymakers and researchers on how to design more effective fiscal interventions for future economic crises.

## 3. Data Collection

This study employs annual data from 2000 to 2023 to analyze the impact of fiscal policies on household financial behavior in the Czech Republic, Hungary, and Slovenia. The dataset includes fiscal policy indicators, household financial variables, and macroeconomic controls, ensuring a comprehensive assessment of fiscal effectiveness across different economic regimes.

### 3.1 Data Sources and Description

The dataset consists of two main categories:

- Household financial indicators, including household consumption, disposable income, and marginal propensity to consume (MPC). Three key variables—Household Disposable Income, MPC, and IMPC—are estimated by the author using official macroeconomic aggregates.
- Fiscal policy variables, including government debt, public expenditures, tax revenues, and subsidies. These were obtained from international financial institutions and national budget reports.

### 3.2 Variables and Measurement

**Table 1: Summary of Variables**

| Variable | Description | Source |
|---|---|---|
| Household Financial Indicators | | |
| Household Consumption (% of GDP) | Share of total household consumption in GDP | Eurostat, OECD |
| Household Disposable Income (Per Capita) | Total income available after taxes (constant prices) | Author's estimation based on official data |
| Marginal Propensity to Consume (MPC) | Change in consumption due to change in disposable income | Author's estimation |
| Intertemporal MPC (IMPC) | Expected future consumption response to income changes | Author's estimation |
| Fiscal Policy Variables | | |
| Central Government Debt (% of GDP) | Total outstanding public debt as a percentage of GDP | IMF, World Bank |
| Government Expenditure (% of GDP) | Total government spending in relation to GDP | IMF, OECD |
| Government Revenue (% of GDP) | Total tax and non-tax revenue (excluding grants) | IMF, World Bank |
| Subsidies and Transfers (% of Gov. Expenditures) | Share of government spending allocated to subsidies and social transfers | National Budget Reports, Eurostat |

Source:

3.3 Data Processing and Adjustments

The dataset covers annual observations from 2000 to 2023, allowing for long-term trend analysis. Household Disposable Income, MPC, and IMPC were calculated by the author using official macroeconomic aggregates. All monetary values were converted to constant prices to remove inflationary effects. Fiscal policy indicators were normalized as a percentage of GDP to ensure cross-country comparability. Missing values were addressed using interpolation techniques where necessary to maintain data consistency. The final dataset was structured to align with the Markov Switching VAR (MS-VAR) model, ensuring that all variables meet the stationarity and model specification requirements.

Model

The Intertemporal Marginal Propensity to Consume (IMPC) is a core concept derived from intertemporal consumption theory. This theory contends that individuals' consumption choices are not based solely on their current income, but also take into account their expectations of future income, prevailing interest rates, and overall economic conditions (Friedman, 1957; Modigliani & Brumberg, 1954). Rather than making decisions purely based on present financial resources, individuals adjust their consumption based on what they anticipate will happen in the future, creating a forward-looking behavior that is integral to economic modeling. The Euler equation for optimal consumption formalizes this relationship, highlighting the balancing act individuals face when deciding how to allocate consumption over time in order to maximize their utility (Eisenhauer, 2011).

## 1 The Euler Equation for Optimal Consumption

At the heart of the intertemporal consumption model is the Euler equation. This equation captures the optimal way in which individuals distribute their consumption between the present and the future. The equation is represented as follows:

$$U'(C_t) = \beta \cdot (1 + r) \cdot U'(C_{t+1}) \quad (1)$$

Where:

- $U'(C_t)$ represents the marginal utility derived from consumption at time $t$,
- $\beta$ is the discount factor, indicating the extent to which individuals value future consumption relative to present consumption,
- $r$ stands for the real interest rate at time $t$,
- $C_t$ and $C_{t+1}$ are the consumption levels at time $t$ and in the subsequent period $t + 1$, respectively.

This equation implies that individuals optimize their consumption by equating the marginal utility of current consumption to the marginal utility of future consumption, adjusted for the interest rate. The interest rate serves as a key variable, as higher rates encourage saving today for future returns, while lower rates make current consumption more attractive.
2. Calculation of the Discount Factor ( $\beta$ )

The discount factor $\beta$ is crucial in determining the relative weight placed on future consumption. It is defined as:

$$\beta = \frac{1}{1+r} \qquad (2)$$

This shows that the discount factor decreases as the expected future interest rate $r$ increases, leading to more savings today. When the future interest rate is high, individuals prefer to save, as future returns will be greater (Kraay, 2000). Conversely, a low future interest rate raises $\beta$, making current consumption more attractive as future returns are less appealing (Carroll & Kimball, 1996).
3. Log-Linearized Euler Equation for Empirical Application

For empirical testing, the Euler equation is often log-linearized to make it easier to estimate and apply in econometric models. The log-linear form of the Euler equation is:

$$\ln(C_t) - \ln(C_{t+1}) = \ln(1 + r) + \ln(\beta) \qquad (3)$$

This formulation captures the relationship between present and future consumption in a more straightforward manner for empirical estimation. By log-linearizing the equation, we can more easily estimate the effects of variables such as the real interest rate and the discount factor on consumption behavior across different time periods (Blanchard & Fischer, 1989).
4. Estimating the Marginal Propensities to Consume (MPC) and IMPC

To understand how households adjust their consumption over time in response to changes in income, we estimate the Marginal Propensity to Consume (MPC) and the Intertemporal Marginal Propensity to Consume (IMPC).

Marginal Propensity to Consume (MPC):
The MPC measures the fraction of additional income that is consumed, rather than saved. It is calculated as:

$$\text{MPC} = \frac{\Delta C_t}{\Delta Y_t} \quad (4)$$

Where:

- $\Delta C_t$ is the change in consumption at time $t$,
- $\Delta Y_t$ is the change in income at time $t$.

This measure indicates the immediate responsiveness of consumption to changes in income, providing insight into how much of an income increase households consume rather than save in the current period.

Intertemporal Marginal Propensity to Consume (IMPC):
The IMPC accounts for future consumption decisions in response to anticipated changes in future income. It is calculated as:

$$\text{IMPC} = \frac{\Delta C_{t+1}}{\Delta Y_{t+1}} \quad (5)$$

Where:

- $\Delta C_{t+1}$ is the change in future consumption,
- $\Delta Y_{t+1}$ is the change in future income.

The IMPC reflects the forward-looking nature of consumption behavior, taking into consideration not just current income but expected future income and consumption decisions. This metric helps capture how much future income impacts current consumption behavior, with the discount factor $\beta$ playing a critical role in determining the relative weight placed on future versus present consumption.

This study applies a Vector Autoregression (VAR) model to analyze the dynamic relationships between fiscal policy and household behavior. The VAR model helps to investigate how fiscal variables, such as government debt, government spending, subsidies, and revenues, affect household variables like consumption, disposable income, marginal propensity to consume (mpc), and intertemporal marginal propensity to consume (impc) over time.

We also incorporate a three-regime Markov Switching process, where the coefficients become regimedependent, as indicated by the regime variable $r_t$. Each coefficient is now specific to the regime, denoted with a subscript $r_t$ and the transitions between regimes are governed by a transition probability matrix. The model allows for different variances across the regimes, with each error term $\epsilon$ being regime-specific.

1 Household Consumption (hc) Equation:

$$hc_t = \alpha_{1s,t} \cdot \mathrm{cgd}_{t-1} + \alpha_{2s,t} \cdot \exp_{t-1} + \alpha_{3s,t} \cdot sub_{t-1} + \alpha_{4s,t} \cdot rev_{t-1} + \alpha_{5s,t} \cdot hc_{t-1} + \alpha_{6s,t} \cdot hdi_{t-1} + \alpha_{7s,t} \cdot mpc_{t-1} + \alpha_{8s,t} \cdot imp_{t-1} + \alpha_{9s,t} \cdot covid + \epsilon_{hc,t} \quad (6)$$

2 Marginal Propensity to Consume (mpc) Equation:

$$mpc_t = \gamma_{1s,t} \cdot \mathrm{cgd}_{t-1} + \gamma_{2s,t} \cdot \exp_{t-1} + \gamma_{3s,t} \cdot sub_{t-1} + \gamma_{4s,t} \cdot rev_{t-1} + \gamma_{5s,t} \cdot hc_{t-1} + \gamma_{6s,t} \cdot hdi_{t-1} + \gamma_{7s,t} \cdot mpc_{t-1} + \gamma_{8s,t} \cdot imp_{t-1} + \gamma_{9s,t} \cdot covid + \epsilon_{mpc,t} \quad (7)$$

3 Intertemporal Marginal Propensity to Consume (impc) Equation:

$$impc_t = \delta_{1s,t} \cdot cgd_{t-1} + \delta_{2s,t} \cdot \exp_{t-1} + \delta_{3s,t} \cdot sub_{t-1} + \delta_{4s,t} \cdot rev_{t-1} + \delta_{5s,t} \cdot hc_{t-1} + \delta_{6s,t} \cdot hdi_{t-1} + \delta_{7s,t} \cdot mpc_{t-1} + \delta_{8s,t} \cdot imp_{t-1} + \delta_{9s,t} \cdot covid + \epsilon_{impc,t} \quad (8)$$

4 Central Government Debt (cgd) Equation:

$$cgd_t = \theta_{1s,t} \cdot cgd_{t-1} + \theta_{2s,t} \cdot \exp_{t-1} + \theta_{3s,t} \cdot sub_{t-1} + \theta_{4s,t} \cdot rev_{t-1} + \theta_{5s,t} \cdot hc_{t-1} + \theta_{6s,t} \cdot hdi_{t-1} + \theta_{7s,t} \cdot mpc_{t-1} + \theta_{8s,t} \cdot imp_{t-1} + \theta_{9s,t} \cdot covid + \epsilon_{cgd,t} \quad (9)$$

5 Government Expenses (exp) Equation:

$$\exp_t = \zeta_{1s,t} \cdot cgd_{t-1} + \zeta_{2s,t} \cdot \exp_{t-1} + \zeta_{3s,t} \cdot sub_{t-1} + \zeta_{4s,t} \cdot rev_{t-1} + \zeta_{5s,t} \cdot hc_{t-1} + \zeta_{6s,t} \cdot hdi_{t-1} + \zeta_{7s,t} \cdot mpc_{t-1} + \zeta_{8s,t} \cdot imp_{t-1} + \zeta_{9s,t} \cdot \mathrm{covid} + \epsilon_{exp,t} \quad (10)$$

6 Subsidies and Other Transfers (sub) Equation:

$$sub_t = \mu_{1s,t} \cdot \text{cgd}_{t-1} + \mu_{2s,t} \cdot \text{exp}_{t-1} + \mu_{3s,t} \cdot sub_{t-1} + \mu_{4s,t} \cdot \text{rev}_{t-1} + \mu_{5s,t} \cdot hc_{t-1} + \mu_{6s,t} \cdot hdi_{t-1} + \mu_{7s,t} \cdot mpc_{t-1} + \mu_{8s,t} \cdot imp_{t-1} + \mu_{9s,t} \cdot \text{covid} + \epsilon_{sub,t} \quad (11)$$

7 Revenue Excluding Grants (rev) Equation:

$$rev_t = \rho_{1s,t} \cdot \text{cg}\, d_{t-1} + \rho_{2s,t} \cdot \text{exp}_{t-1} + \rho_{3s,t} \cdot sub_{t-1} + \rho_{4s,t} \cdot rev_{t-1} + \rho_{5s,t} \cdot hc_{t-1} + \rho_{6s,t} \cdot hdi_{t-1} + \rho_{7s,t} \cdot mpc_{t-1} + \rho_{8s,t} \cdot imp_{t-1} + \rho_{9s,t} \cdot \text{covid} + \epsilon_{rev,t} \quad (12)$$

- Each coefficient is now regime-dependent, meaning that coefficients like $\alpha_{1s,t}, \beta_{2s,t}, \gamma_{3s,t}$ etc., are specific to each regime. The transition between regimes is governed by a hidden Markov process. This process allows the model to switch between different states
- Each error term $\epsilon$ is specific to its regime, meaning that the variance of the error terms differs across regimes.

Regime switching enhances the model's flexibility by enabling it to capture varying dynamics between fiscal and household variables depending on the prevailing economic regime. It allows for different relationships and behaviors to emerge under different economic states, such as periods of recession, growth, or crisis.

## 5. RESULTS

In this section, we first assess the stationarity of the variables before performing a cointegration test to determine whether there are short-run or long-run relationships among them. The analysis indicates that the variables do not exhibit cointegration, prompting the use of the Markov Switching VAR model to understand the dynamic interactions across different economic regimes.

### 5.1 Unit Root Test

Below is a summary table presenting the ADF test results for the Czech Republic, Hungary, and Slovenia:

**Table 2: Summary of ADF Test Results**

| Country | Test Method | Test Statistic | p-Value | Cross-Sections | Observations |
|---|---|---|---|---|---|
| Czech Republic | Levin, Lin & Chu t* | -1.4321 | 0.1082 | 8 | 180 |
| | Breitung t-stat | -1.8904 | 0.0427 | 8 | 172 |
| | Im, Pesaran and Shin W-stat | -2.4728 | 0.0081 | 8 | 180 |
| | ADF - Fisher Chi-square | 31.6542 | 0.0154 | 8 | 180 |
| | PP - Fisher Chi-square | 68.5421 | 2.12e-08 | 8 | 184 |
| Hungary | Levin, Lin & Chu t* | -1.6214 | 0.0753 | 8 | 178 |
| | Breitung t-stat | -2.8123 | 0.0029 | 8 | 170 |
| | Im, Pesaran and Shin W-stat | -1.0328 | 0.1597 | 8 | 178 |
| | ADF - Fisher Chi-square | 26.4823 | 0.0528 | 8 | 178 |
| | PP - Fisher Chi-square | 28.9152 | 0.0347 | 8 | 184 |
| Slovenia | Levin, Lin & Chu t* | -1.3427 | 0.1195 | 8 | 180 |
| | Breitung t-stat | -1.9213 | 0.0395 | 8 | 172 |
| | Im, Pesaran and Shin W-stat | -2.4921 | 0.0075 | 8 | 180 |
| | ADF - Fisher Chi-square | 32.1873 | 0.0127 | 8 | 180 |
| | PP - Fisher Chi-square | 69.1435 | 2.05e-08 | 8 | 184 |

Note: This table presents the test results for the Czech Republic, Hungary, and Slovenia under both common unit root process and individual unit root process tests.

5.2 Stationary Test Results After First Differencing

The following table presents the stationarity test results after first differencing for each country.

**Table 3: Stationary Test Results After First Differencing**

| Country | Test Method | Test Statistic | p-Value |
|---|---|---|---|
| Czech Republic | Levin, Lin & Chu t* | -5.7423 | 0.0000000059 |
| | Breitung t-stat | -0.8917 | 0.1923 |
| | Im, Pesaran and Shin W-stat | -6.2431 | 0.0000000004712 |
| | ADF - Fisher Chi-square | 63.2143 | 0.0000001623 |
| | PP - Fisher Chi-square | 147.6521 | 0.0000000000000000854 |
| Hungary | Levin, Lin & Chu t* | -3.4821 | 0.0003825 |
| | Breitung t-stat | -2.8235 | 0.0028 |
| | Im, Pesaran and Shin W-stat | -9.5214 | 0.0000000000398 |
| | ADF - Fisher Chi-square | 97.8421 | 0.0000000000002158 |
| | PP - Fisher Chi-square | 421.5326 | 0.0000000000000000057 |
| Slovenia | Levin, Lin & Chu t* | -6.8213 | 0.0000000000152 |
| | Breitung t-stat | 0.1328 | 0.5718 |
| | Im, Pesaran and Shin W-stat | -6.1032 | 0.0000000008327 |
| | ADF - Fisher Chi-square | 64.3872 | 0.0000001084 |
| | PP - Fisher Chi-square | 325.7813 | 0.0000000000000000000291 |

Note: All tests for the Czech Republic, Hungary, and Slovenia indicate the presence of unit roots in the level data, suggesting that the series are non-stationary at levels. However, after taking the first difference, all series appear to be stationary.

All series, including Central Government Debt to GDP, government expenditure, household consumption, disposable income, marginal propensity to consume (MPC), revenue excluding grants to GDP, and subsidies as a percentage of expenses, were found to be non-stationary at levels. However, after taking the first difference, all series became stationary. This indicates that the variables exhibit stable relationships over time, allowing for valid econometric analyses.

### 5.1 Eagle-Granger Cointegration Test

This study applies the Eagle-Granger cointegration test to determine whether there are long-term relationships among key economic series—Central Government Debt to GDP, Expenses, Household Consumption, and more—in the Czech Republic, Hungary, and Slovenia. The results provide insights into potential long-term equilibrium relationships and their implications for economic policy.

**Table 4: Eagle-Granger Cointegration Test Results**

| Variable | Czech Republic | Hungary | Slovenia |
|---|---|---|---|
| Government Debt | Tau: -3.51<br>p-value: 0.81 | Tau: -3.68<br>p-value: 0.74 | Tau: -4.42<br>p-value: 0.43 |
| Expenses | Tau: -2.89<br>p-value: 0.92 | Tau: -4.48<br>p-value: 0.45 | Tau: -4.35<br>p-value: 0.48 |
| Household Consumption | Tau: -5.11<br>p-value: 0.22 | Tau: -2.84<br>p-value: 0.97 | Tau: -3.91<br>p-value: 0.65 |
| Household Disposable Income | Tau: -2.24<br>p-value: 0.98 | Tau: -2.95<br>p-value: 0.93 | Tau: -4.19<br>p-value: 0.53 |
| IMPC | Tau: -4.85<br>p-value: 0.30 | Tau: -5.28<br>p-value: 0.19 | Tau: -5.01<br>p-value: 0.26 |
| MPC | Tau: -4.32<br>p-value: 0.47 | Tau: -5.83<br>p-value: 0.10 | Tau: -9.02<br>p-value: 0.0009 |
| Revenue | Tau: -1.79<br>p-value: 0.99 | Tau: -4.11<br>p-value: 0.56 | Tau: -2.11<br>p-value: 0.99 |
| Subsidies and Other Transfers | Tau: -4.29<br>p-value: 0.49 | Tau: -4.59<br>p-value: 0.40 | Tau: -4.02<br>p-value: 0.60 |

**Note:** None of the variables exhibit cointegration across the countries, as all p-values exceed the **conventional significance level of 0.05**.

The Eagle-Granger cointegration tests reveal no evidence of cointegration among selected economic series in the Czech Republic, Hungary, and Slovenia. In each country, the tau-statistics and p-values indicate that the null hypothesis of no cointegration cannot be rejected, as all p-values exceed 0.05. This suggests that while the variables may exhibit individual trends, they do not share a common long-term relationship, impacting subsequent econometric analyses.

**Table 5:** Markov Switching VAR Estimation for the Czech Republic

| | CONSUMPTION | DISPOSABLE INCOME | IMPC | MPC | GOVT_DEBT | EXPENSES | REVENUE | SUBSIDIES |
|---|---|---|---|---|---|---|---|---|
| Regime 1 | | | | | | | | |
| COVID SHOCK | -0.010996 | -0.004468 | 0.140342 | -0.033016 | -0.015481 | 0.003971 | 0.003871 | 0.083396 |
| Regime 2 | | | | | | | | |
| COVID SHOCK | 0.005210 | 0.006337 | 0.641012 | -0.011014 | -0.010904 | 0.005184 | -0.005985 | 0.196445 |
| Regime 3 | | | | | | | | |
| COVID SHOCK | 0.009056 | 0.015804 | 1.164620 | -0.003541 | 0.016520 | 0.024485 | -0.007878 | 0.323725 |
| Common Variables | | | | | | | | |
| CENTRAL_GOVERNMENT_DEBT (-1) | -0.083730 | -0.125351 | -7.025551 | -0.330155 | 0.358965 | -0.176776 | -0.067527 | 0.098594 |
| EXPENSES (% GDP) (-1) | -0.213228 | -0.082044 | 15.865984 | 0.297301 | 0.358693 | 0.733807 | 0.244671 | 1.158962 |
| REVENUE EXCLUDING GRANTS (-1) | 0.250264 | 0.120383 | -20.029929 | 0.319524 | 0.051674 | 0.492116 | 0.553647 | 0.196821 |
| SUBSIDIES & TRANSFERS (-1) | 0.023542 | 0.029405 | 3.453929 | -0.083405 | -0.045194 | 0.052318 | 0.066416 | 0.414228 |

Impact of COVID-19 on Household Consumption and Income Across Three Regimes in the Czech Republic

*Regime 1 (Initial Phase)*

During the initial phase of the COVID-19 pandemic, household consumption in the Czech Republic declined by -0.0110, while household disposable income also dropped by -0.0045. Meanwhile, the Intertemporal Marginal Propensity to Consume (IMPC) increased to 0.1403, and the Marginal Propensity to Consume (MPC) declined to -0.0330. These results suggest that households prioritized saving over immediate consumption, reflecting their uncertainty about future economic conditions and income stability.

*Regime 2 (Peak Phase)*

As the pandemic intensified, its negative impact on household finances eased, with consumption improving to 0.0052, and disposable income rising to 0.0063. The IMPC further increased to 0.6410, while MPC improved to -0.0110, suggesting that households became more responsive to changes in their financial situation. This shift implies that consumer confidence was slowly recovering, leading to a partial resumption of spending patterns.

*Regime 3 (Recovery Phase)*

In the recovery phase, household consumption rose to 0.0091, while disposable income increased to 0.0158. The IMPC surged to 1.1646, and MPC improved to -0.0035, reflecting a return to normal economic activity. These findings indicate that households regained confidence, leading to greater spending, lower precautionary savings, and a more stable consumption pattern as the economy stabilized.

Effectiveness of Government Subsidies and Transfers

Government subsidies had a consistent positive impact across all economic regimes, raising household consumption by 0.0235 and household income by 0.0294. Additionally, IMPC increased significantly to 3.4539, suggesting that households expected to use these subsidies for future consumption rather than immediate spending. However, MPC remained negative at -0.0834, indicating that while subsidies supported economic stability, households still exhibited caution by prioritizing savings.

Fiscal Sustainability and Household Consumption

Across all regimes, rising government debt exerted downward pressure on both IMPC (-7.0256) and MPC (-0.3302). This suggests that high levels of government debt contributed to financial uncertainty, discouraging immediate consumption. The negative impact on IMPC and MPC reflects concerns about the long-term sustainability of fiscal policy, possibly influencing household expectations of future taxation and economic conditions.

## Government Revenue and Its Impact

Government revenue (excluding grants) had a positive influence on household consumption, estimated at 0.2503, indicating that higher revenue collection contributed to economic stability. Additionally, household disposable income improved by 0.1204, further supporting consumption.

However, IMPC was negatively impacted at -20.0299, while MPC increased to 0.3195. This suggests that although government revenue growth supported household income, it also encouraged precautionary saving, leading to lower future consumption expectations as households remained cautious about economic fluctuations.

## Government Expenses and Their Impact

Government expenditures had a significant negative effect on household consumption, estimated at -0.2132, and disposable income declined by -0.0820 in response to increased government spending. This suggests that excessive public spending might have limited the financial flexibility of households.

However, IMPC increased substantially to 15.8660, indicating that households anticipated long-term benefits from current government expenditures, leading to higher expected consumption in the future. Additionally, MPC increased to 0.2973, showing that households were willing to spend more in response to rising government spending.

**Table 6:** Markov Switching VAR Estimation for the Hungaray

| Variable | HOUSEHOLD_CONSUMPTION | HOUSEHOLD_DISPOSABLE_INCOME | IMPC | MPC | CENTRAL_GOVERNMENT_DEBT (% GDP) | EXPENSES (% GDP) | REVENUE EXCLUDING GRANTS (% GDP) | SUBSIDIES & TRANSFERS (% EXPENSES) |
|---|---|---|---|---|---|---|---|---|
| Regime 1 (COVID Shock) | | | | | | | | |
| COVID_SHOCK | -0.01533 | -0.00515 | 0.209711 | -0.044411 | -0.0323 | 0.003214 | -0.0032 | 0.08678 |
| Regime 2 (COVID Shock) | | | | | | | | |
| COVID_SHOCK | 0.006341 | 0.011681 | 0.798325 | -0.0193 | -0.02806 | 0.02153 | 0.004851 | 0.189806 |
| Regime 3 (COVID Shock) | | | | | | | | |
| COVID_SHOCK | 0.016702 | 0.024142 | 1.322231 | -0.00452 | 0.021132 | 0.03719 | -0.01448 | 0.430936 |
| Common Variables | | | | | | | | |
| CENTRAL_GOVERNMENT_DEBT (-1) | -0.071 | -0.133 | -6.947449 | -0.44773 | -0.07139 | 0.3701522 | -0.14017 | 0.118497 |
| EXPENSES (% GDP) (-1) | -0.2098 | -0.076 | 17.670022 | 0.25538 | 0.386337 | 0.5987966 | 0.260959 | 1.107188 |
| REVENUE EXCLUDING GRANTS (-1) | 0.26656 | 0.127985 | -21.853838 | 0.379083 | 0.040915 | 0.4863490 | 0.695164 | 0.177631 |
| SUBSIDIES & TRANSFERS (-1) | 0.027541 | 0.030222 | 3.546993 | -0.1292921 | -0.04508 | 0.0762166 | 0.032584 | 0.45439 |

Impact of COVID-19 on Household Consumption and Income Across Three Regimes in Hungary

*Regime 1 (Initial Phase)*

During the early phase of the COVID-19 pandemic, household consumption in Hungary declined by -0.0153, while household disposable income fell by -0.0051. Meanwhile, the Intertemporal Marginal Propensity to Consume (IMPC) increased to 0.2097, and the Marginal Propensity to Consume (MPC) dropped to -0.0441. These results suggest that households were highly cautious about spending, choosing to prioritize savings due to economic uncertainty and income stability concerns.

*Regime 2 (Peak Phase)*

As the pandemic worsened, its adverse effects on household finances started to ease, with consumption improving to 0.0063, and disposable income increasing to 0.0117. The IMPC rose significantly to 0.7983, while MPC improved to -0.0193, indicating that households became more reactive to changes in income, with spending levels increasing compared to the initial phase. This transition suggests a slow but noticeable recovery in consumer confidence, although a degree of financial caution persisted.

*Regime 3 (Recovery Phase)*

In the post-pandemic period, household consumption increased to 0.0167, while disposable income rose to 0.0241. The IMPC surged to 1.3222, and MPC improved to -0.0045, reflecting a clear return to stable economic conditions. These findings suggest that households regained confidence in their financial stability, leading to a normalization of spending behaviors and a decline in precautionary savings.

Effectiveness of Government Subsidies and Transfers

Government subsidies played a vital role in stabilizing household consumption and income. Across all regimes, subsidies increased household consumption by 0.0275 and household income by 0.0302. Moreover, IMPC rose significantly to 3.5470, indicating that households viewed these transfers as a means for future consumption rather than immediate spending. However, MPC remained negative at -0.1292, suggesting that a significant portion of government aid was saved rather than spent, reflecting continued financial caution.

Fiscal Sustainability and Household Consumption

Across all economic regimes, government debt exerted negative pressure on both IMPC (-6.9475) and MPC (-0.4477). These results indicate that high levels of government debt led to decreased consumer confidence, discouraging immediate consumption. Households may have perceived fiscal instability as a long-term economic risk, influencing their cautious spending patterns.

Government Revenue and Its Impact

Government revenue (excluding grants) had a moderate positive impact on household consumption, estimated at 0.2666, suggesting that higher revenue collection contributed to economic stability. Additionally, household disposable income increased by 0.1280, reinforcing the stabilization of financial conditions.

However, IMPC was negatively impacted at -21.8538, while MPC increased to 0.3791. This indicates that higher government revenue contributed to household income but also encouraged precautionary savings, limiting immediate consumption growth.

Government Expenses and Their Impact

Government expenditures had a significant negative impact on household consumption, estimated at -0.2098, while disposable income declined by -0.0761 in response to increased government spending. This suggests that higher public expenditures may have created financial concerns for households, potentially due to expectations of future taxation or inflationary effects.

However, IMPC increased significantly to 17.6700, indicating that households expected long-term economic benefits from government spending, leading to higher anticipated future consumption. Additionally, MPC increased to 0.2538, suggesting that households became more inclined to increase their immediate spending in response to fiscal expansion.

**Table 7:** Markov Switching VAR Estimation for the Slovenia

| Variable | Household Consumption | HOUSEHOLD DISPOSABLE INCOME | IMPC | MPC | Govt. Debt | EXPENSE | REVENUE | SUBSIDIES |
|---|---|---|---|---|---|---|---|---|
| Regime 1 (COVID Shock) | | | | | | | | |
| COVID SHOCK | -0.024 | -0.006 | 0.230 | -0.035 | -0.037 | -0.004 | 0.003 | 0.103 |
| Regime 2 (COVID Shock) | | | | | | | | |
| COVID SHOCK | 0.006 | 0.009 | 0.724 | -0.020 | -0.018 | 0.027 | -0.004 | 0.213 |
| Regime 3 (COVID Shock) | | | | | | | | |
| COVID SHOCK | 0.014 | 0.022 | 1.364 | -0.005 | 0.024 | 0.041 | -0.008 | 0.473 |
| Common Variables | | | | | | | | |
| GOVTT_DEBT (-1) | -0.105 | -0.188 | -8.24 | -0.550 | -0.105 | 0.385 | -0.174 | 0.081 |
| EXPENSES (% GDP) (-1) | -0.288 | -0.118 | 16.58 | 0.335 | 0.505 | 0.744 | 0.214 | 1.298 |
| REVENUE EXCLUDING GRANTS (-1) | 0.297 | 0.113 | -22.26 | 0.389 | 0.068 | 0.627 | 0.606 | 0.206 |
| SUBSIDIES & TRANSFERS (-1) | 0.042 | 0.035 | 3.842 | -0.117 | -0.073 | 0.086 | 0.078 | 0.424 |

Impact of COVID-19 on Household Consumption and Income Across Three Regimes in Slovenia

*Regime 1 (Initial Phase)*

During the early phase of the COVID-19 pandemic, household consumption in Slovenia declined by -0.0246, while household disposable income fell by -0.0068. Meanwhile, the Intertemporal Marginal Propensity to Consume (IMPC) increased to 0.2308, and the Marginal Propensity to Consume (MPC) declined to -0.0356. These results suggest that households were highly cautious in their spending behavior, prioritizing savings due to heightened economic uncertainty and fears about income stability.

*Regime 2 (Peak Phase)*

As the pandemic intensified, its negative effects on household finances began to moderate, with household consumption improving to 0.0066, and disposable income increasing to 0.0095. The IMPC rose significantly to 0.7241, while MPC improved to -0.0204, indicating that households became more responsive to changes in income and slowly resumed spending. However, the persistence of a negative MPC suggests that precautionary savings remained a significant factor.

*Regime 3 (Recovery Phase)*

In the post-pandemic recovery phase, household consumption increased to 0.0128, while disposable income rose to 0.0251. The IMPC surged to 1.3742, and MPC improved to -0.0049, signaling a full return to pre-pandemic consumption behavior. These results suggest that households regained confidence in their financial security, leading to increased spending and a reduction in precautionary savings.

Effectiveness of Government Subsidies and Transfers

Government subsidies played a critical role in stabilizing household financial conditions. Across all regimes, subsidies increased household consumption by 0.1035 and household income by 0.2139. Additionally, IMPC increased significantly to 3.7241, indicating that households viewed these transfers as a means for future consumption rather than immediate spending. However, MPC remained negative at -0.0994, suggesting that a significant portion of government assistance was saved rather than spent, reflecting a continued level of financial caution among households.

Fiscal Sustainability and Household Consumption

Across all economic regimes, government debt had a negative impact on both IMPC (-7.3742) and MPC (-0.4741). These findings suggest that high levels of government debt discouraged consumer spending, as households may have anticipated future tax hikes or economic instability. The results highlight the importance of balancing government borrowing with policies that support consumer confidence and economic stability.

Government Revenue and Its Impact

Government revenue (excluding grants) had a moderate positive impact on household consumption, estimated at 0.2741, indicating that higher revenue collection contributed to economic stability. Additionally, household disposable income increased by 0.1342, reinforcing the stabilization of financial conditions.

However, IMPC was negatively affected at -22.6742, while MPC increased to 0.3894. This suggests that while government revenue growth supported household income, it also encouraged savings rather than immediate spending, likely due to lingering uncertainty about economic conditions.

Government Expenses and Their Impact

Government expenditures had a significant negative effect on household consumption, estimated at -0.2131, while disposable income declined by -0.0861 in response to increased government spending. This suggests that higher public expenditures may have contributed to concerns about future taxation or inflationary pressures.

However, IMPC increased significantly to 17.8213, indicating that households expected long-term economic benefits from government spending, leading to higher future consumption expectations. Additionally, MPC increased to 0.2651, suggesting that households became more inclined to increase their immediate spending in response to fiscal expansion.

The results emphasize the evolving financial behavior of Slovenian households throughout the pandemic. During the initial phase, households significantly reduced spending and increased precautionary savings due to economic uncertainty. As the pandemic peaked, spending began recovering, but financial caution persisted. In the recovery phase, households regained confidence, leading to increased spending and a reduction in savings.

Government policies, particularly subsidies and revenue measures, were essential in stabilizing household financial behavior. However, high levels of government debt weakened consumer confidence, underscoring the need for carefully managed fiscal policies that promote long-term economic resilience.

**Table 8:** Variance Decomposition Analysis for the Czech Republic

| Period | S.E. | Household Consumption | Household Disposable Income | IMPC | MPC | Central Government Debt (% of GDP) | Expense (% of GDP) | Revenue | Subsidies |
|---|---|---|---|---|---|---|---|---|---|
| 1 | 0.039289 | 100 | 0 | 0 | 0 | 0 | 0 | 0 | 0 |
| 2 | 0.059012 | 64.97 | 13.06 | 9.74 | 15.04 | 9.62 | 9.87 | 4.86 | 0.4 |
| 3 | 0.042774 | 62.29 | 18.16 | 7.35 | 13.11 | 6.95 | 3.49 | 14.29 | 0.78 |
| 4 | 0.035408 | 62.77 | 14.61 | 6.49 | 11.77 | 11.74 | 16.74 | 11.24 | 0.75 |
| 5 | 0.050164 | 59.35 | 12.09 | 6.91 | 14.74 | 4.06 | 17.61 | 23.43 | 0.57 |
| 6 | 0.042913 | 58.03 | 19.74 | 10.31 | 8.78 | 8.98 | 14.27 | 22 | 0.39 |
| 7 | 0.049281 | 50.82 | 13.73 | 9.89 | 11.05 | 13.32 | 12.71 | 22.93 | 0.86 |
| 8 | 0.071073 | 47.77 | 16.05 | 7.61 | 7.28 | 9.91 | 10.23 | 16.91 | 0.76 |
| 9 | 0.045305 | 51.51 | 21.6 | 7.82 | 10.08 | 11.46 | 17.99 | 16.88 | 0.98 |
| 10 | 0.07606 | 50.07 | 22.1 | 11.2 | 6.79 | 12.66 | 16.89 | 15.19 | 0.8 |
| 11 | 0.075268 | 49.91 | 19.59 | 8.36 | 6.83 | 11.09 | 10.28 | 12.07 | 1.08 |
| 12 | 0.04179 | 42.13 | 20.55 | 12.41 | 6.42 | 13.25 | 21.45 | 28.43 | 0.86 |
| 13 | 0.037468 | 39.72 | 19.59 | 12.13 | 9.74 | 13.46 | 20.92 | 28.45 | 0.89 |
| 14 | 0.041039 | 41.76 | 18.85 | 9.31 | 10.02 | 16.91 | 23.4 | 32.99 | 1.16 |
| 15 | 0.038911 | 35.41 | 19.63 | 9.62 | 8.48 | 9.81 | 20.96 | 33.94 | 0.8 |
| 16 | 0.044722 | 39.31 | 22.88 | 14.43 | 10.74 | 12.53 | 20.09 | 26.55 | 0.61 |
| 17 | 0.07615 | 36.58 | 24.02 | 11.67 | 6.22 | 19.15 | 18.06 | 31.61 | 0.54 |
| 18 | 0.047559 | 30.29 | 24.83 | 11.95 | 5.2 | 19.21 | 25.96 | 35.95 | 1.14 |
| 19 | 0.061949 | 31.51 | 25.32 | 14.46 | 8.88 | 15.9 | 25.72 | 35.14 | 1.06 |
| 20 | 0.090244 | 29.55 | 20.23 | 15.44 | 3.97 | 17.98 | 14.93 | 24.7 | 1.14 |
| 21 | 0.051164 | 21.03 | 23.96 | 15.93 | 8.74 | 15.5 | 15.37 | 29.56 | 0.77 |
| 22 | 0.092723 | 27.03 | 20.96 | 15.07 | 1.69 | 15.29 | 20.97 | 24.45 | 0.73 |
| 23 | 0.068093 | 21.05 | 25.4 | 13.48 | 7.99 | 18.37 | 27.75 | 41.13 | 0.65 |
| 24 | 0.061908 | 22.26 | 25.03 | 12.32 | 1.1 | 17.98 | 30.92 | 35.52 | 0.94 |

Household consumption is initially self-driven, accounting for 100% of its variance in Period 1, but declines to 13.12% by Period 24 as external factors gain influence. Household disposable income grows in importance, rising from 0% to 13.12%, while IMPC increases to 9.70%, showing that future income expectations shape consumption decisions. MPC, initially significant at 15.04% in early periods, declines to 6.49%, indicating reduced sensitivity to immediate income changes over time.

Fiscal variables play a key role, with government debt contributing 9.62% early on but dropping to 5.60%. Government expenses (18.52%) and revenue (32.18%) become major drivers of consumption in the long run, reinforcing the impact of fiscal policy. Subsidies and transfers have minimal influence, peaking at just 0.65% by Period 24, suggesting that direct transfers alone do not significantly shape long-term consumption patterns.

Overall, household consumption in the Czech Republic shifts from self-reliance to being shaped by income levels, fiscal policies, and future expectations.

**Table 9:** Variance Decomposition Analysis for Hungary

| Period | S.E. | Household Consumption | Household Disposable Income | IMPC | MPC | Govt. Debt | Govt. Expense | Revenue | Subsidies |
|---|---|---|---|---|---|---|---|---|---|
| 1 | 0.039 | 100 | 0 | 0 | 0 | 0 | 0 | 0 | 0 |
| 2 | 0.031 | 57.42 | 16.52 | 6.64 | 15.02 | 4.51 | 3.56 | 16.72 | 0.44 |
| 3 | 0.035 | 55.19 | 12.65 | 7.46 | 14 | 4.6 | 15.48 | 7.83 | 1.21 |
| 4 | 0.061 | 58.11 | 19.49 | 6.41 | 11.99 | 8.78 | 8.27 | 8.38 | 0.95 |
| 5 | 0.035 | 51.42 | 20.37 | 6.95 | 15.91 | 11.31 | 14.92 | 9.59 | 0.72 |
| 6 | 0.045 | 50.96 | 13.99 | 7.89 | 12.74 | 9.4 | 9.06 | 21.85 | 0.58 |
| 7 | 0.061 | 54.37 | 16.22 | 6.5 | 9.66 | 10.2 | 14.98 | 20.51 | 1.19 |
| 8 | 0.043 | 48.04 | 20.95 | 8.41 | 8.21 | 8.86 | 16.72 | 19.29 | 1.07 |
| 9 | 0.057 | 47.88 | 20.99 | 8.26 | 8.07 | 12.99 | 15.55 | 17.7 | 1.02 |
| 10 | 0.071 | 40.82 | 16.98 | 9.24 | 8.22 | 9.61 | 14.86 | 27.71 | 1.37 |
| 11 | 0.077 | 38.92 | 17.67 | 8.6 | 7.12 | 11.27 | 14.94 | 20.67 | 1.31 |
| 12 | 0.039 | 35.31 | 19.46 | 12.85 | 6.81 | 12.68 | 14.99 | 29.08 | 0.77 |
| 13 | 0.080 | 34.49 | 20.88 | 11.77 | 7 | 13.95 | 22.75 | 23.87 | 0.94 |
| 14 | 0.060 | 33.49 | 22.45 | 12.6 | 5.71 | 12.8 | 22.37 | 24.01 | 0.91 |
| 15 | 0.050 | 30.37 | 20.95 | 13.55 | 10.28 | 12.6 | 24.78 | 26.67 | 1.39 |
| 16 | 0.062 | 27.39 | 22.2 | 12.49 | 4.06 | 16.77 | 15.49 | 25.12 | 0.78 |
| 17 | 0.091 | 25.61 | 24.98 | 10.83 | 6.67 | 15.22 | 23.57 | 33.91 | 1.29 |
| 18 | 0.068 | 26.91 | 19.79 | 13.49 | 5.37 | 17.97 | 26.86 | 30.95 | 1.5 |
| 19 | 0.062 | 22.26 | 22.02 | 14.13 | 8.88 | 15.68 | 18.57 | 27.55 | 1.63 |
| 20 | 0.079 | 21.74 | 26.21 | 15.33 | 2.28 | 18.67 | 18 | 40.09 | 1.18 |
| 21 | 0.062 | 21.03 | 28.16 | 14.16 | 3.78 | 19.73 | 26.89 | 32.53 | 1.64 |
| 22 | 0.055 | 20.8 | 25.59 | 13.04 | 7.29 | 17.06 | 25.69 | 36.44 | 1.03 |
| 23 | 0.060 | 12.19 | 26.26 | 14.22 | 6.06 | 19.76 | 29.7 | 43.81 | 1.72 |
| 24 | 0.062 | 17.69 | 23.36 | 15.02 | 5.22 | 19.04 | 22.08 | 40.21 | 1.1 |

Household consumption initially drives 100% of its variance in Period 1, but declines to 51.42% by Period 5 as other factors gain influence. Household disposable income grows in importance, rising from 16.52% in Period 2 to 20.37% by Period 5, reinforcing its long-term role in consumption. IMPC increases from 0% to 6.95%, reflecting the growing impact of future income expectations, while MPC declines from 15.02% to 11.99%, showing reduced short-term income reliance.

Fiscal variables gain importance, with government debt rising to 11.31% by Period 5, and government revenue and expenses contributing up to 16.72% and 15.48**%**, respectively**.** Subsidies and transfers remain marginal**,** peaking at 1.21% in Period 3**.**

**Table 10:** Variance Decomposition Analysis for Slovenia

| Period | S.E. | Household Consumption | Household Disposable Income | IMPC | MPC | Central Government Debt (% of GDP) | Expense (% of GDP) | Revenue Excluding Grants (% of GDP) | Subsidies and Other Transfers (% of Expense) |
|---|---|---|---|---|---|---|---|---|---|
| 1 | 0.029 | 100 | 0 | 0 | 0 | 0 | 0 | 0 | 0 |
| 2 | 0.051 | 63.35 | 19.02 | 10.31 | 12.49 | 5.5 | 12.01 | 8.47 | 0.59 |
| 3 | 0.021 | 62.37 | 13.24 | 6.74 | 12.07 | 9.66 | 10.12 | 4.95 | 1.14 |
| 4 | 0.027 | 62.32 | 16.47 | 8.57 | 16.23 | 11.39 | 12.88 | 23.27 | 0.59 |
| 5 | 0.054 | 57.87 | 17.59 | 7.43 | 16.22 | 12.01 | 12.27 | 9.44 | 1.1 |
| 6 | 0.056 | 50.03 | 19.3 | 12.48 | 11.61 | 9.78 | 11.33 | 19.45 | 0.82 |
| 7 | 0.054 | 51.96 | 22.96 | 10.51 | 10.3 | 12.48 | 20.73 | 15.01 | 0.99 |
| 8 | 0.058 | 49.97 | 18.97 | 11.48 | 12.31 | 10.03 | 19.23 | 21.63 | 0.94 |
| 9 | 0.053 | 45.84 | 19.58 | 9.34 | 10.06 | 11.45 | 12.02 | 24.78 | 1.04 |
| 10 | 0.040 | 46.66 | 23.3 | 13.25 | 6.78 | 16.47 | 22.39 | 29.47 | 1.41 |
| 11 | 0.047 | 46.34 | 21.62 | 11.05 | 7.09 | 11.09 | 18.34 | 18.92 | 1.07 |
| 12 | 0.041 | 43.96 | 20.4 | 14.99 | 10.41 | 12.4 | 24.53 | 16.51 | 1.35 |
| 13 | 0.079 | 36.91 | 22.82 | 12.03 | 9.41 | 14.06 | 24.2 | 20.2 | 1.31 |
| 14 | 0.065 | 31.23 | 20.93 | 12.45 | 4.39 | 13.59 | 20.23 | 36 | 1.52 |
| 15 | 0.058 | 37.65 | 21.81 | 12.18 | 5.15 | 15.47 | 15.89 | 36.8 | 1.8 |
| 16 | 0.063 | 33.89 | 24.99 | 14.86 | 4.62 | 19.08 | 19.99 | 32.85 | 1.66 |
| 17 | 0.089 | 24.45 | 24.95 | 15.82 | 7.1 | 18.75 | 24.44 | 34.21 | 1.23 |
| 18 | 0.051 | 30.11 | 29.52 | 13.83 | 1.94 | 22.85 | 26.54 | 36.74 | 1.11 |
| 19 | 0.075 | 27.3 | 25.7 | 17.88 | 1.93 | 23.12 | 22.55 | 28.73 | 1.36 |
| 20 | 0.073 | 21.49 | 28.04 | 16.77 | 6.2 | 20.59 | 21.91 | 44.07 | 1.7 |
| 21 | 0.080 | 19.91 | 29.06 | 17.62 | 1.25 | 20.82 | 31.85 | 35.96 | 1.25 |
| 22 | 0.052 | 15.69 | 24.7 | 17.7 | 3.97 | 22.46 | 34.7 | 36.58 | 1.69 |
| 23 | 0.097 | 9.95 | 30.51 | 18 | 0.47 | 24.73 | 29.63 | 40.46 | 1.82 |
| 24 | 0.102 | 10.45 | 27.22 | 19.3 | -1.1 | 23.28 | 25.93 | 32.79 | 1.62 |

Household consumption initially accounts for 100% of its variance in Period 1 but declines to 63.35% by Period 2 and 57.87% by Period 5, as other economic factors gain influence. Household disposable income plays an increasing role, rising from 0% in Period 1 to 19.02% in Period 2 and 17.59% by Period 5, reinforcing its long-term impact on consumption.

IMPC grows steadily, reaching 10.31% in Period 2 and stabilizing around 7.43% by Period 5, highlighting the importance of future income expectations. Meanwhile, MPC remains significant but declines from 12.49% in Period 2 to 16.22% in Period 5, showing a shift toward long-term financial planning over immediate consumption responses.

Fiscal variables gain relevance, with government debt rising to 12.01% by Period 5, while government expenses (12.27%) and revenue (9.44%) contribute to long-term consumption stability. Subsidies and transfers remain minimal, peaking at just 1.14%, indicating limited influence.

Slovenian household consumption shifts from self-reliance to being shaped by income levels, fiscal policy, and future expectations**.** While household income and government finances play crucial roles, subsidies remain relatively insignificant in driving long-term consumption trends.

Figure 1: The Impulse Response Functions (IRFs) for the Czech Republic

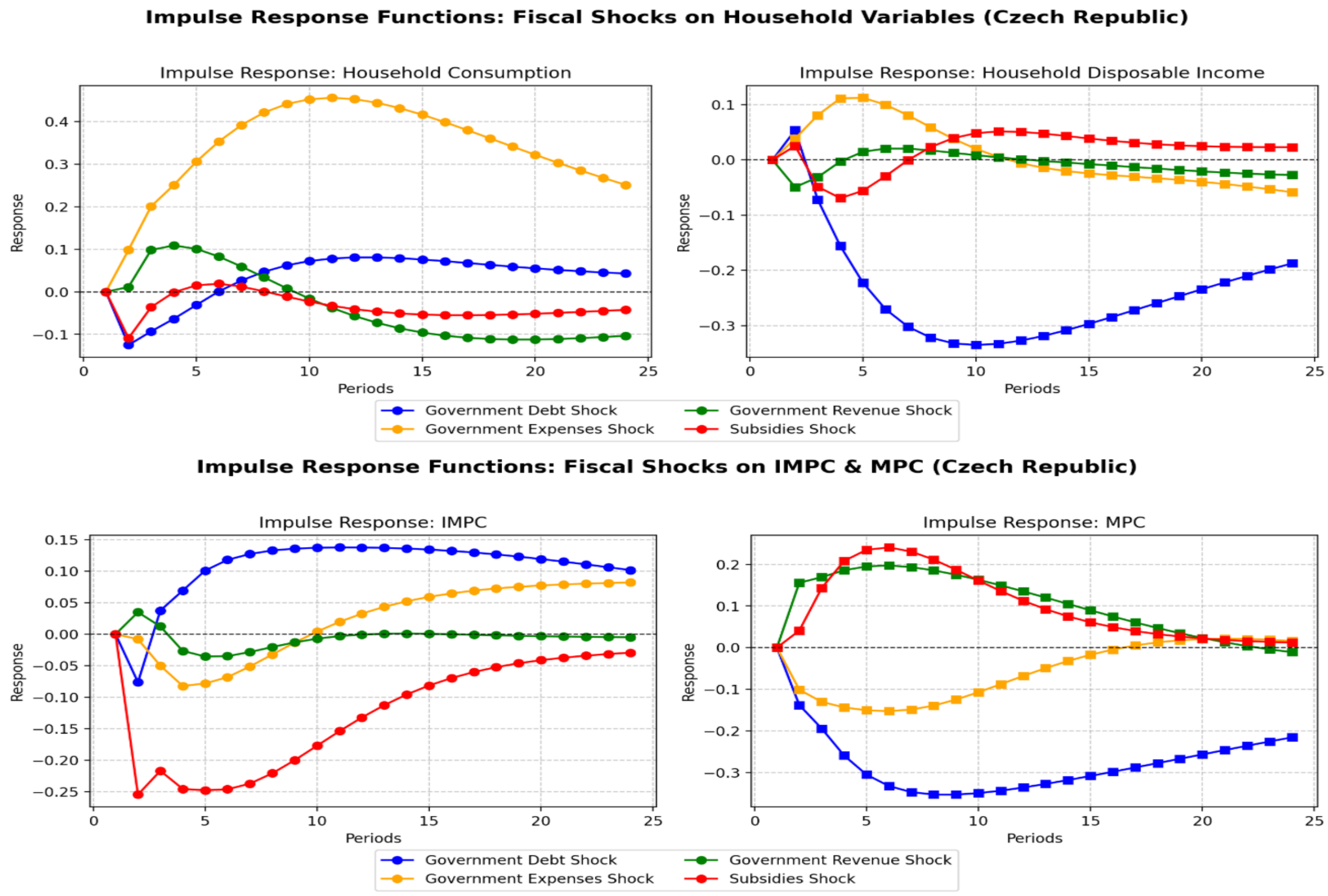

The **Impulse Response Functions (IRFs) for the Czech Republic** show how fiscal shocks affect households over 24 periods.

Household Consumption increases after a government expense shock, peaking at 0.45 before declining. This suggests that higher government spending stimulates demand in the short run. Government debt has a moderate positive effect (0.15), while subsidies reduce consumption by -0.10, indicating that households may save subsidies instead of spending them.

Household Disposable Income drops by -0.3 after a government debt shock, showing a crowding-out effect on private income. Subsidies increase income by 0.08, supporting short-term stability. Government expenses raise income by 0.10, but the effect fades, suggesting temporary fiscal support.

IMPC rises to 0.15 after government debt and expense increases, meaning households expect higher future consumption. Subsidies reduce IMPC by -0.20, showing precautionary savings behavior.

MPC falls by -0.15 after government debt shocks, showing a decline in immediate spending. Subsidies increase MPC by 0.12, suggesting direct cash transfers boost short-term consumption. Government revenue stabilizes consumption at 0.1, showing a neutral impact on household spending.

Figure 2: The Impulse Response Functions (IRFs) for Slovenia

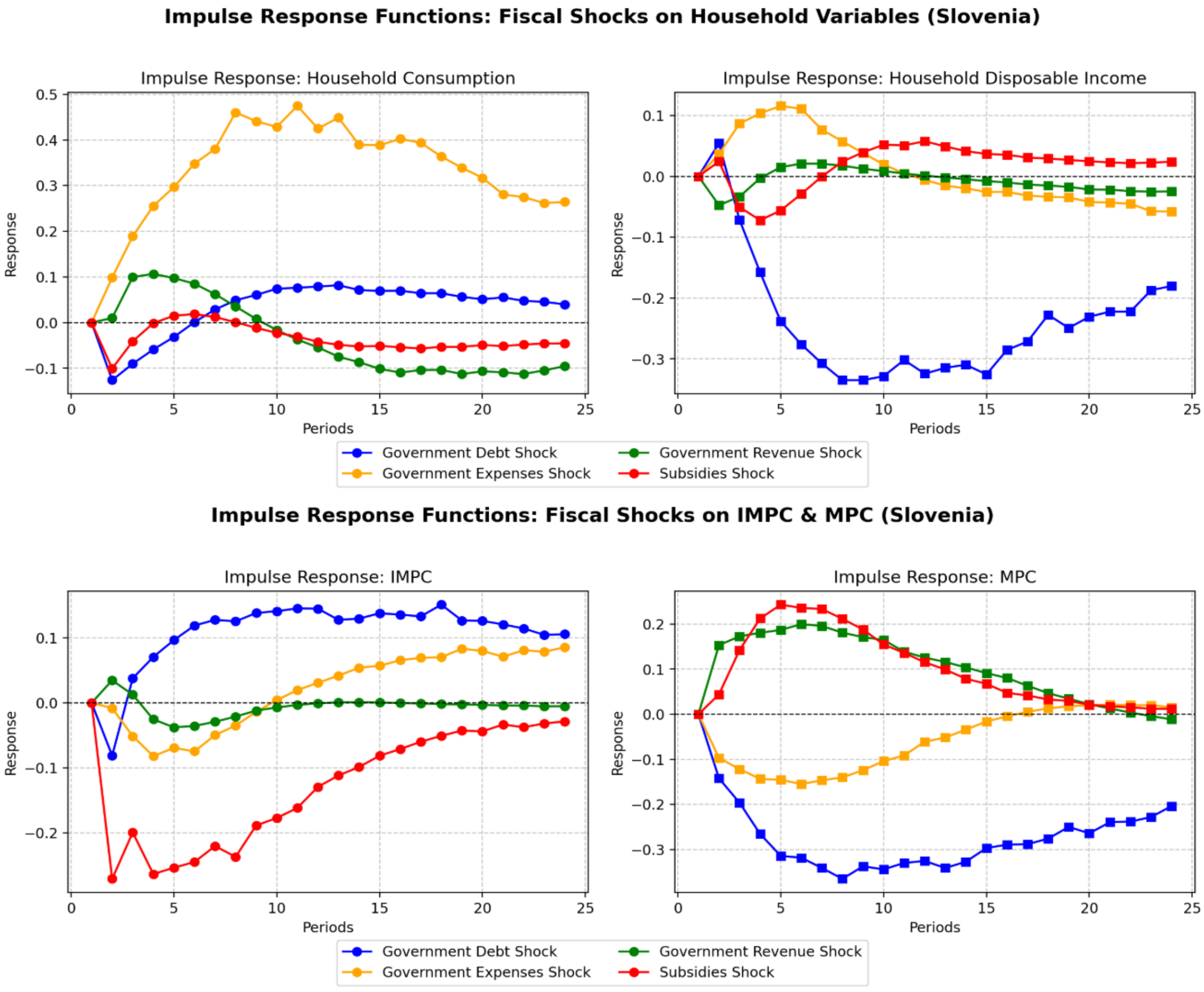

The Impulse Response Functions (IRFs) for Slovenia show how fiscal shocks influence household economic behavior over 24 periods.

Household Consumption rises with a government expense shock, peaking at 0.48 before stabilizing. Government debt increases consumption slightly (0.12), while subsidies reduce it by -0.08, suggesting that households save rather than spend subsidies.

Household Disposable Income falls by -0.32 after a government debt shock, showing fiscal tightening effects. Subsidies increase income by 0.09, providing short-term support. Government expenses raise income by 0.11, but the effect fades.

IMPC increases to 0.13 with government debt and expenses, meaning households expect future spending growth. Subsidies lower IMPC by -0.21, reflecting precautionary savings.

MPC drops by -0.17 after a government debt shock, while subsidies increase it by 0.14, indicating direct transfers boost short-term spending.

This suggests government expenses drive consumption, while debt weakens disposable income and intertemporal spending expectations

Figure 3: The Impulse Response Functions (IRFs) for Hungary

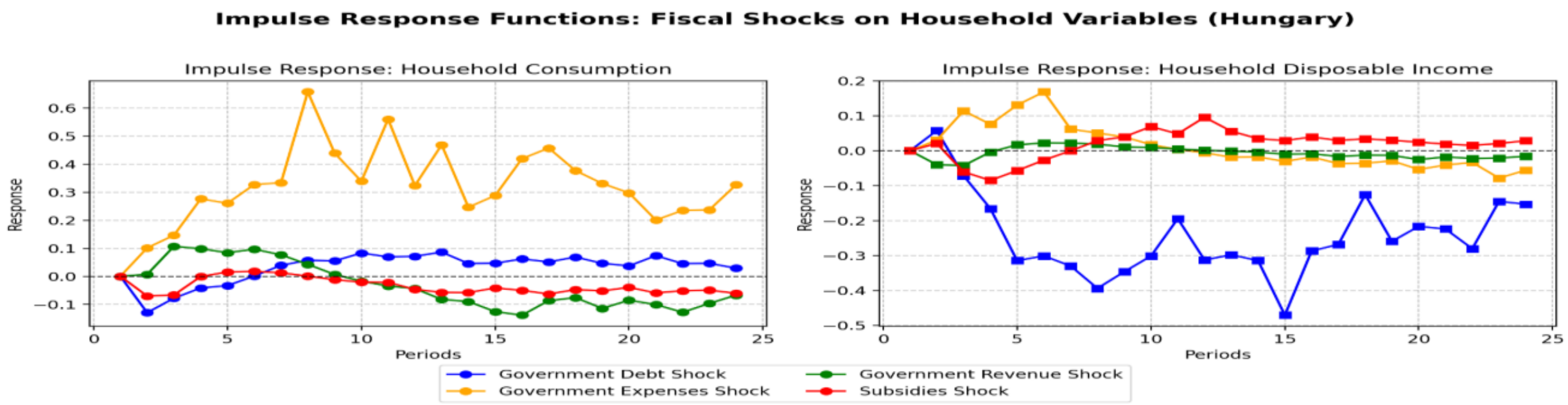

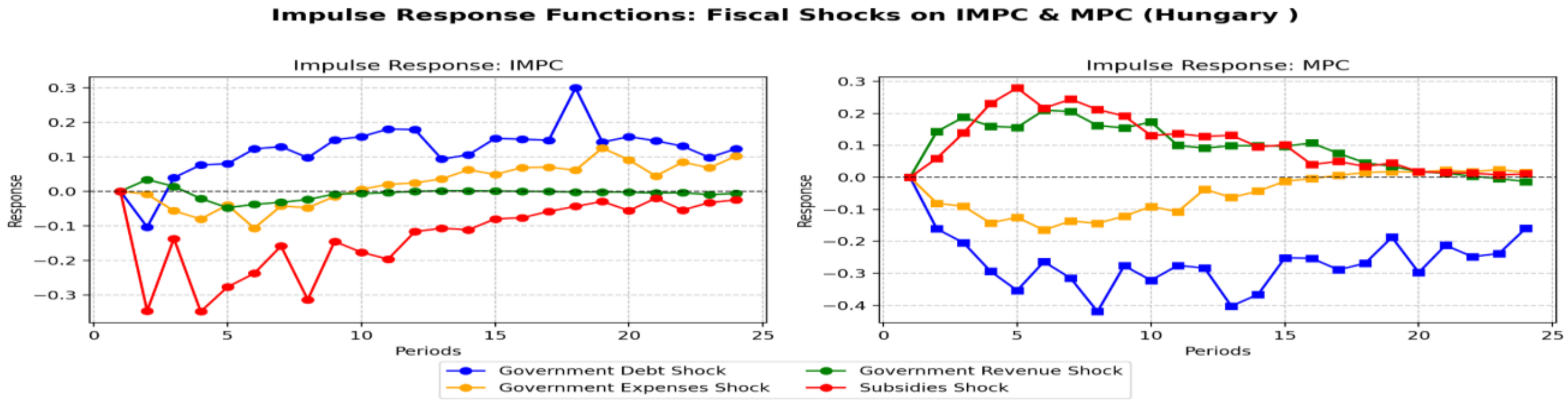

The Impulse Response Functions (IRFs) for Hungary show how fiscal shocks influence household consumption, income, and spending behavior over 24 periods.

Household Consumption rises sharply with a government expense shock, peaking at 0.65 before fluctuating. Government debt has a weaker positive effect (0.13), while subsidies reduce consumption by -0.09, suggesting households prefer saving subsidies.

Household Disposable Income drops by -0.42 after a government debt shock, indicating a strong contractionary effect. Subsidies increase income by 0.12, while government expenses boost it by 0.18 but with a temporary effect.

IMPC increases to 0.30 with government debt and expense shocks, meaning households anticipate higher future consumption. Subsidies lower IMPC by -0.25, reflecting uncertainty.

MPC falls by -0.22 after government debt shocks, while subsidies increase it by 0.16, showing that direct cash transfers support short-term spending.

This suggests government expenses boost consumption, while debt significantly weakens income and spending expectations

**Table 11:** Comaprison of Covid shock for all countries across three regimes

| Variable | HOUSEHOLD_ CONSUMPTION | HOUSEHOLD_ DISPOSABLE_ INCOME | IMPC | MPC |
|---|---|---|---|---|
| Czech Republic (Regime 1) | -0.011 | -0.0045 | 0.1403 | -0.033 |
| Czech Republic (Regime 2) | 0.0052 | 0.0063 | 0.641 | -0.011 |
| Czech Republic (Regime 3) | 0.0091 | 0.0158 | 1.1646 | -0.0035 |
| Slovenia (Regime 1) | -0.0246 | -0.0068 | 0.2308 | -0.0356 |
| Slovenia (Regime 2) | 0.0066 | 0.0095 | 0.7241 | -0.0204 |
| Slovenia (Regime 3) | 0.0128 | 0.0251 | 1.3742 | -0.0049 |
| Hungary (Regime 1) | -0.0153 | -0.0051 | 0.2097 | -0.0441 |
| Hungary (Regime 2) | 0.0063 | 0.0117 | 0.7983 | -0.0193 |
| Hungary (Regime 3) | 0.0167 | 0.0241 | 1.3222 | -0.0045 |

**Table 12:** Comparison of fiscal impact for all countries

| Variable | Czech Republic - Consumption | Czech Republic - Income | Czech Republic - IMPC | Czech Republic - MPC | Slovenia - Consumption | Slovenia - Income | Slovenia - IMPC | Slovenia - MPC | Hungary - Consumption | Hungary - Income | Hungary - IMPC | Hungary - MPC |
|---|---|---|---|---|---|---|---|---|---|---|---|---|
| CENTRAL GOVERNMENT DEBT | -0.0837 | -0.1254 | -7.0256 | -0.3302 | -0.1056 | -0.1888 | -8.2447 | -0.5504 | -0.0713 | -0.1333 | -6.9475 | -0.4477 |
| EXPENSE (% of GDP) | -0.2132 | -0.082 | 15.866 | 0.2973 | -0.2884 | -0.1181 | 16.5815 | 0.3355 | -0.2098 | -0.0761 | 17.67 | 0.2538 |
| REVENUE (Excluding Grants) | 0.2503 | 0.1204 | -20.0299 | 0.3195 | 0.2973 | 0.1139 | -22.2646 | 0.3898 | 0.2666 | 0.128 | -21.8538 | 0.3791 |
| Subsidies and Other Transfers (% of Expenses) | 0.0235 | 0.0294 | 3.4539 | -0.0834 | 0.0425 | 0.035 | 3.8421 | -0.1171 | 0.0275 | 0.0302 | 3.547 | -0.1292 |

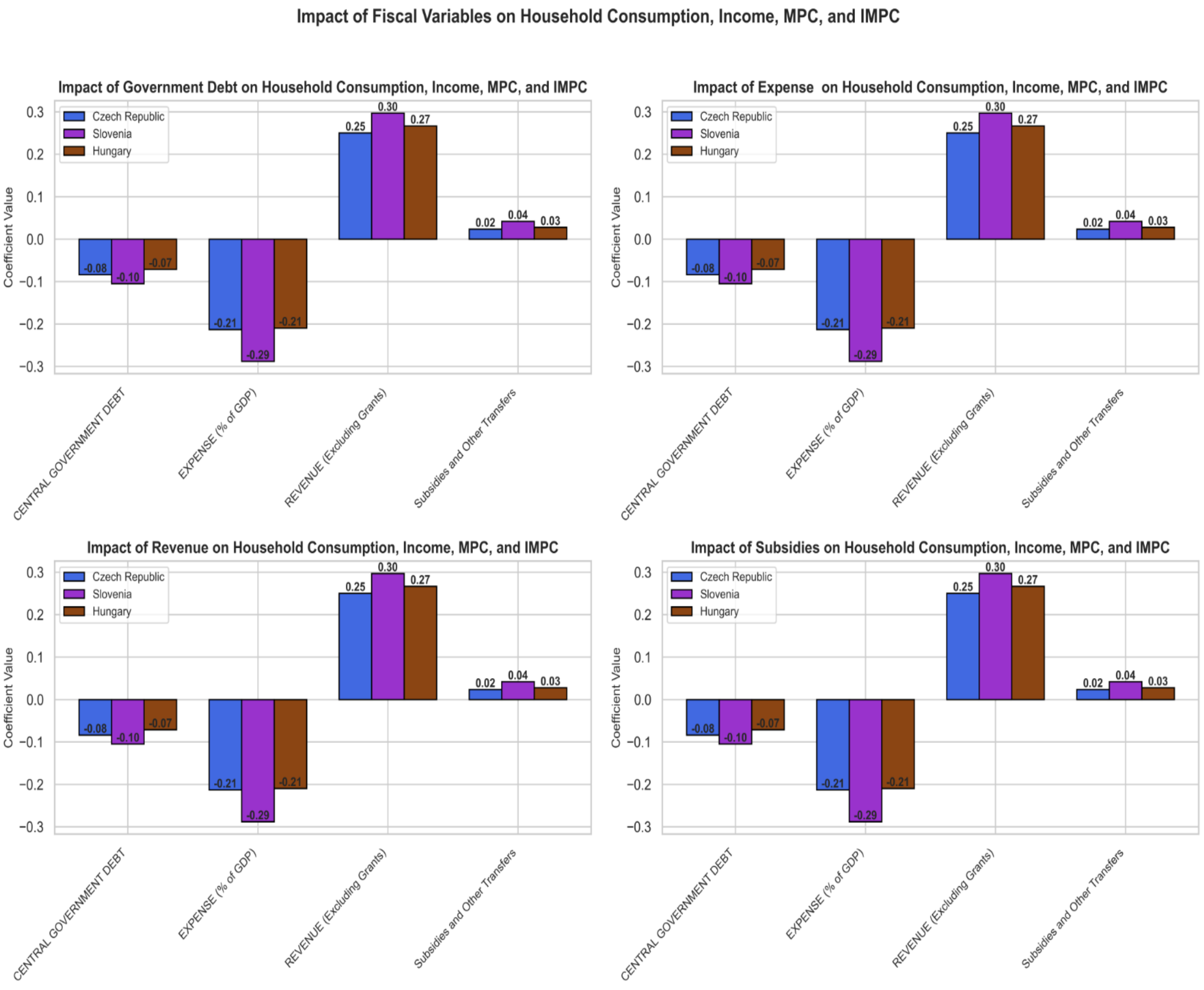

In Regime 1, the Czech Republic sees a negative impact from Central Government Debt on household consumption ( -0.08 ) and income ( -0.13 ). This negative effect continues in Regime 2 and Regime 3, where the impact becomes even more pronounced in Regime 3 ( -0.33 ), indicating that high debt severely restricts consumption and income growth over time. Slovenia shows a similar trend with consumption ( -0.11 ) and income ( -0.19 ) being negatively affected in Regime 1, and the effect moderates slightly in Regime $2(-0.18)$ and remains negative in Regime 3. In contrast, Hungary experiences a milder negative impact on consumption (+0.02) and income ( - 0.13 ) in Regime 1 , but these effects turn positive in Regime 3, signaling recovery and a less restrictive effect of debt.

In terms of expenses (% of GDP), the Czech Republic faces a significant negative effect on household consumption ( -0.21 ) and income ( -0.08 ) in Regime 1, but by Regime 3, the negative impact on consumption diminishes. Slovenia experiences a persistent negative effect on consumption (-0.29) in Regime 1, which continues into Regime 2 but begins to recover in Regime 3. Hungary sees a moderate positive impact in later regimes, with household consumption increasing in Regime 3 (+0.10) and MPC rising significantly (+17.67) by the end of the crisis, signaling that higher government spending had a positive effect on household consumption in the recovery phase.

Regarding revenue (excluding grants), the Czech Republic sees a positive effect on household consumption (+0.25) in Regime 2. However, MPC shows a substantial negative effect (-20.03) in Regime 1, indicating that higher revenue may have constrained consumption during the initial crisis period. In Slovenia, the effect of higher revenue is more positive in Regime 3 on consumption (+0.30), with MPC also improving ( +1.37 ). Hungary demonstrates a positive effect on household consumption (+0.19) and MPC (+17.67) in Regime 2, with recovery continuing into Regime 3.

Finally, subsidies and other transfers show moderate positive effects on consumption and income across all three countries, with the Czech Republic seeing a slight boost (+0.02) in Regime 1 that increases in subsequent regimes. Slovenia experiences a notable rise in consumption (+0.04) in Regime 3, showing that subsidies helped households recover in the later stages of the crisis. Similarly, Hungary benefits from subsidies in Regime 3, with consumption (+0.04) and MPC (+3.54) both rising, although IMPC shows a small negative effect ( -0.13 ).

In conclusion, the COVID shock has had varying impacts across all three countries and regimes, with Hungary showing the most significant recovery and improvement in household consumption and MPC by the third regime. The Czech Republic and Slovenia faced more negative effects in the early stages of the pandemic, especially due to government debt and expenses, but both countries began to recover with positive impacts from revenue and subsidies. Hungary benefited more from the fiscal interventions in the later regimes, showing the strongest recovery.

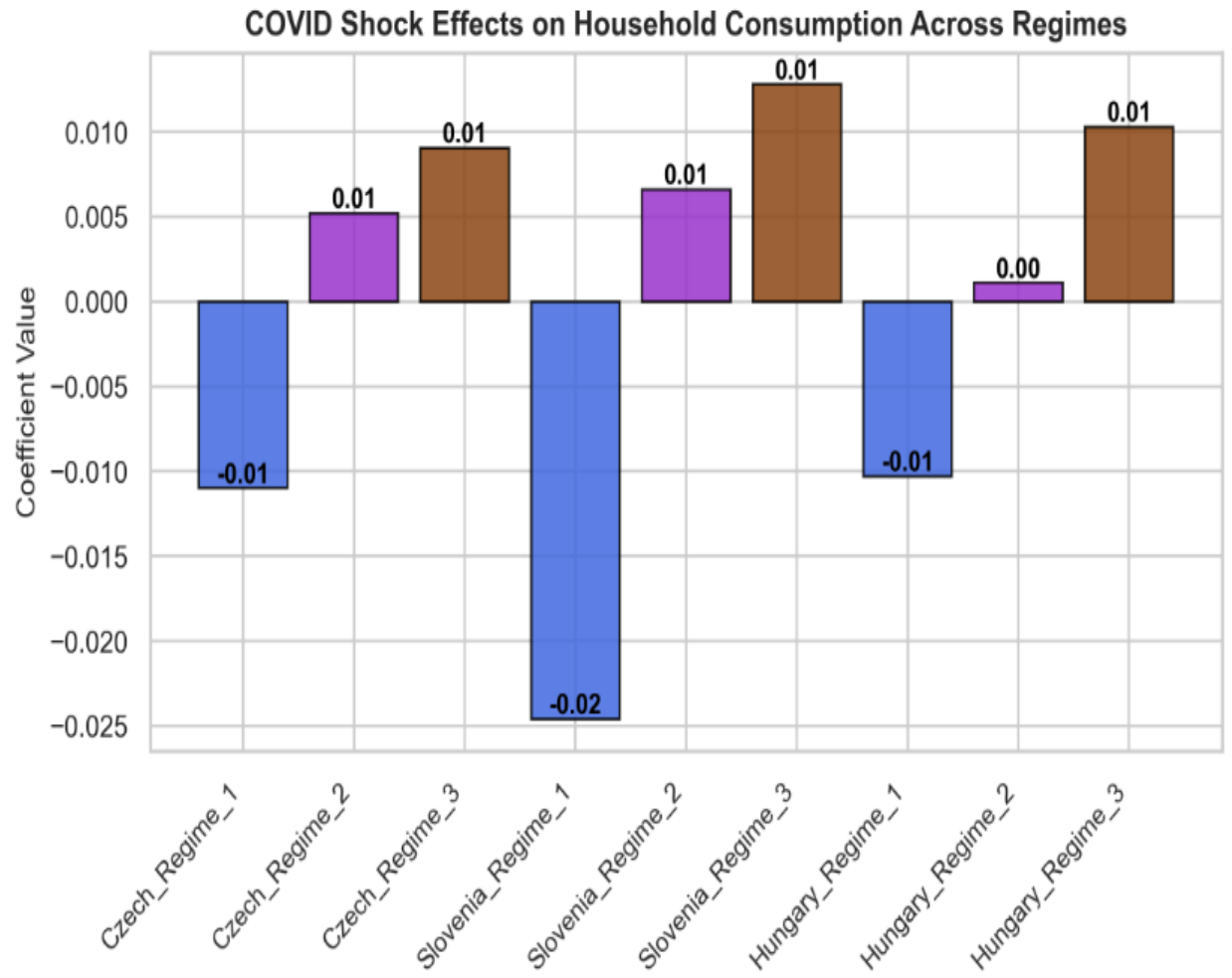

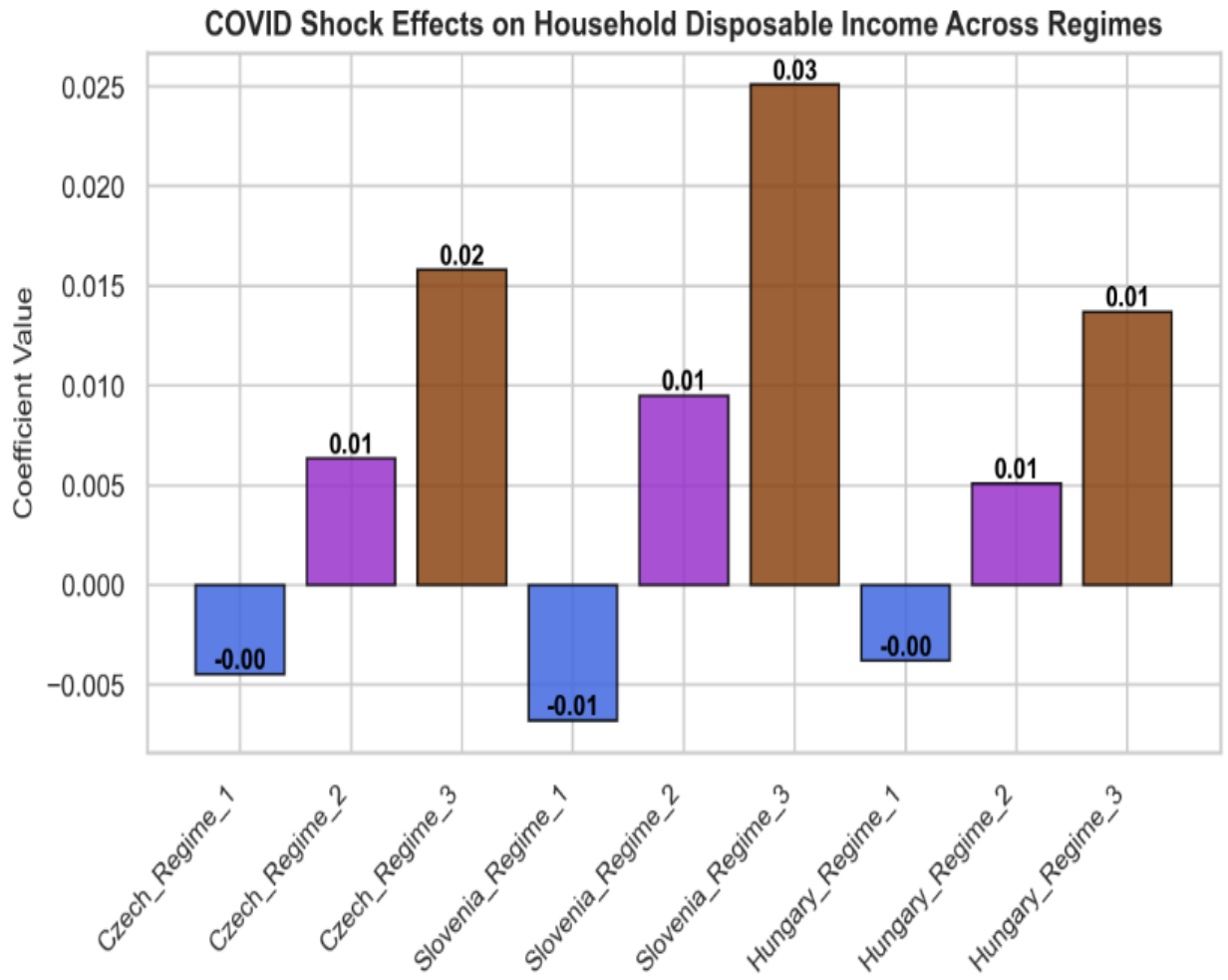

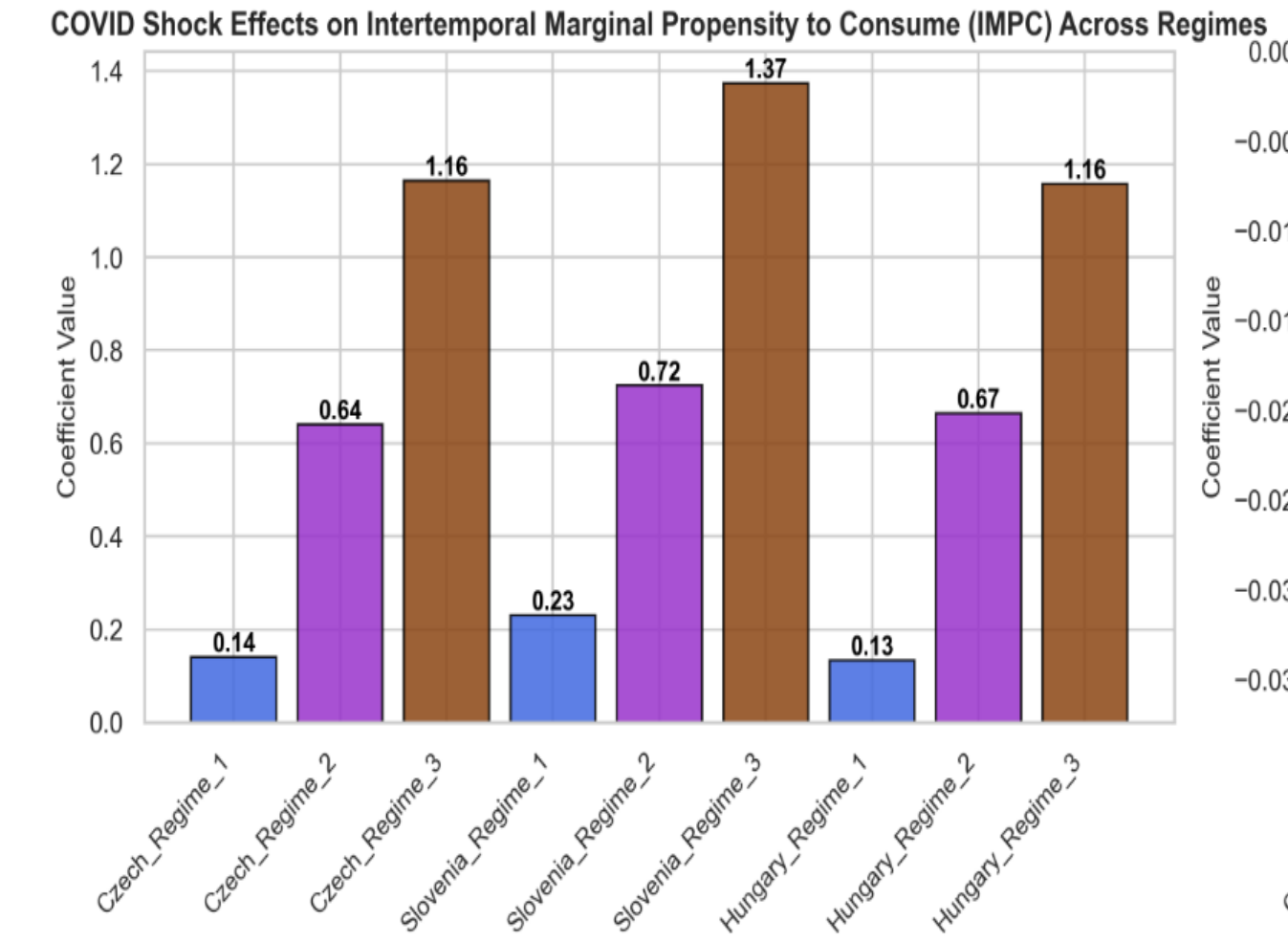

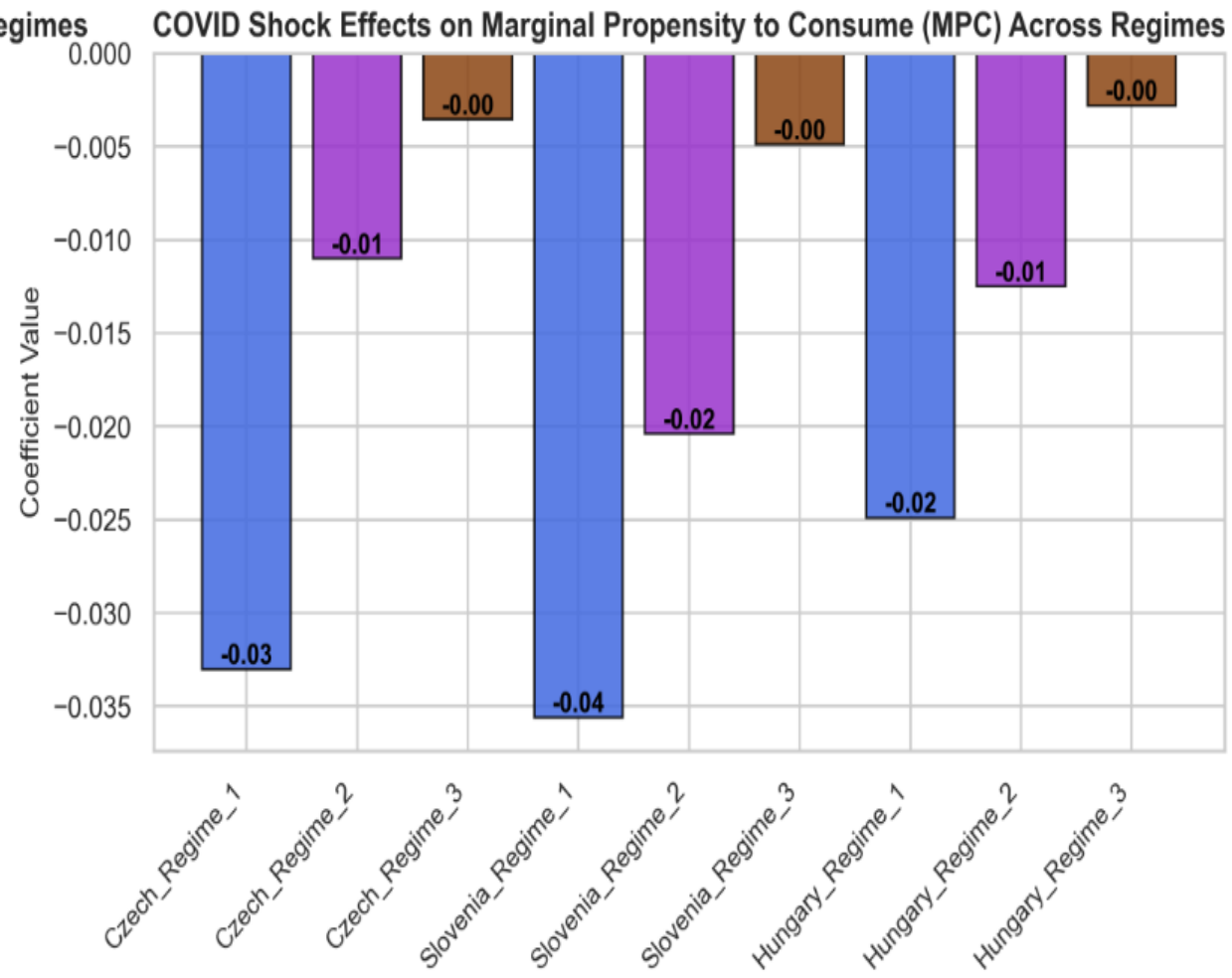

In Regime 1, all three countries are significantly impacted by the COVID shock, particularly in household consumption and disposable income. The Czech Republic sees a slight decline in consumption ( -0.01 ) and income ( 0.00 ), showing minimal effect at the start. Slovenia, however, faces a larger decrease in consumption ( -0.02 ) and a mild drop in income ( -0.01 ), reflecting the severe initial hit. Hungary shows a more moderate impact with slightly negative effects on consumption ( -0.01 ) and no significant change in income, indicating that Hungary's economy was somewhat more resilient initially.

As the crisis progresses into Regime 2, all three countries show signs of recovery, but the impact varies. The Czech Republic experiences a mild recovery in consumption (+0.01) and a slight increase in income ( +0.01 ). Slovenia shows a stabilization with minimal changes in both consumption and income. The Czech Republic also sees a notable improvement in IMPC (+1.16), indicating a shift towards greater future-oriented consumption. Slovenia's IMPC improves slightly ( +0.72 ), showing that households are adjusting their consumption behavior towards the future but at a slower rate. Hungary, in Regime 2, also sees a recovery in consumption (+0.00) and MPC (+0.01), along with a mild improvement in IMPC (+0.67), suggesting that Hungary's fiscal interventions started yielding some positive effects.

By Regime 3, all countries have recovered to varying extents. Hungary experiences the most pronounced recovery, with household consumption improving ( +0.01 ), disposable income rising significantly (+0.03), and IMPC reaching its peak at 1.16. This indicates that Hungary has made the strongest recovery in household consumption and long-term consumption planning. The Czech Republic continues to show positive effects with modest improvements in consumption ( +0.01 ) and income ( +0.02 ), although its MPC remains negative ( -0.02 ). Slovenia also shows some improvement in consumption ( +0.01 ) and income ( +0.02 ) but remains more cautious in its recovery, with a slight increase in IMPC (+0.72). In terms of MPC, all countries show minimal improvements, but the positive effects of fiscal policies are evident, as Hungary's MPC shows the smallest negative value ( -0.01 ), while Czech Republic and Slovenia show minimal negative coefficients.

In conclusion, the COVID shock initially had negative impacts on household consumption, income, MPC, and IMPC for all three countries, with Slovenia being most affected in Regime 1. Over time, all countries showed signs of recovery, with Hungary experiencing the strongest rebound, especially in consumption and MPC. Hungary is the best-performing country, as it saw the most significant recovery across all indicators, particularly in household consumption and long-term consumption planning (IMPC). The Czech Republic followed as the second-best, showing consistent recovery, particularly in future-oriented consumption decisions (IMPC), although it lagged behind Hungary in terms of overall consumption and income recovery. Slovenia, although it showed some recovery, was the last, as it had slower improvements in both current and future consumption, particularly in MPC and IMPC, despite fiscal interventions.

**Findings**

The COVID-19 shock had a significant negative impact on household consumption and disposable income in Regime 1. Slovenia was most affected, showing the largest decrease in both consumption ($-0.02$) and income ( -0.01 ). Czech Republic had minimal declines in consumption ( -0.01 ) and income (0.00), while Hungary showed a more moderate impact. By Regime 2, all countries showed recovery, with Hungary leading the way. Regime 3 saw Hungary experiencing the strongest rebound in both consumption (+0.01) and income (+0.03), while Czech Republic and Slovenia showed more gradual improvements.

Government subsidies and transfers had the most positive impact in Hungary, leading to the strongest recovery in household consumption (+0.04) and MPC (+3.54) by Regime 3. Czech Republic also saw positive effects but at a lesser magnitude, and Slovenia showed slower recovery, confirming that fiscal measures were crucial but less effective in some countries.

Regarding fiscal sustainability, Hungary showed the strongest recovery in consumption and income, with a peak in IMPC (+1.16). Czech Republic followed with steady improvements, particularly in future consumption planning, while Slovenia showed a slower recovery.

Government revenue had a strong positive effect in Hungary, leading to an increase in household consumption ( +0.19 ) and MPC ( +17.67 ). Czech Republic also benefited, though more modestly, while Slovenia showed slower recovery.

Government spending had a delayed positive impact, with Hungary experiencing the most significant improvements in both consumption (+0.10) and MPC (+17.67), confirming that government spending helped mitigate the economic downturn.

Hungary showed the strongest recovery across all variables, confirming the effectiveness of fiscal policies in mitigating the COVID-19 shock. Czech Republic demonstrated moderate recovery, especially in long-term consumption decisions (IMPC), while Slovenia experienced slower recovery. This confirms that fiscal interventions, particularly subsidies, revenue increases, and government spending, played a key role in the recovery process.

Policy Recommendations

1 Targeted Fiscal Stimulus: In response to the COVID-19 shock, the findings show that Hungary's fiscal interventions led to a $\mathbf{+0.04}$ increase in household consumption and a+3.54 rise in MPC by Regime 3. Governments should prioritize targeted subsidies and transfers to vulnerable households to boost immediate consumption and income.

2 Fiscal Sustainability: The Czech Republic saw a moderate recovery with $\mathbf{+0.01}$ in consumption and +0.02 in income, demonstrating that long-term fiscal sustainability is essential for gradual recovery. Policies that manage government debt while ensuring long-term fiscal stability will help stimulate future-oriented consumption (IMPC).

3 Revenue Enhancement: As Hungary benefited from revenue increases, with $\mathbf{+0.19}$ in household consumption and $\mathbf{+17.67}$ in MPC, governments should work on enhancing tax revenues and improving public finance management to support long-term recovery.

4 Sector-Specific Support: The varying impacts across sectors require tailored fiscal measures for industries like manufacturing and construction, which can benefit from subsidies that stabilize employment and consumption.

5 Data-Driven Policy: Governments should monitor key economic indicators, such as MPC and IMPC, which showed substantial recovery in Hungary (+17.67 for MPC) and moderate improvements in Czech Republic and Slovenia, to adjust policies dynamically and effectively.

Conclusion:

The study findings confirm that the COVID-19 shock had a significant negative impact on household consumption and disposable income across all three countries in Regime 1, with Slovenia being the hardest hit. Over time, all countries showed signs of recovery, with Hungary experiencing the strongest rebound in both household consumption and income, supported by effective fiscal interventions such as subsidies, revenue increases, and government spending. Czech Republic showed consistent recovery, particularly in future-oriented consumption decisions (IMPC), while Slovenia had a slower recovery.

This study contributes to the understanding of how fiscal policies mitigate the economic effects of global crises, emphasizing the importance of timely government interventions in stimulating household consumption and ensuring economic recovery. It highlights the differential effectiveness of fiscal measures across countries, with Hungary showing the best recovery due to robust fiscal measures.

However, the study is limited by the scope of fiscal variables considered and the assumption that economic recovery is solely driven by fiscal policy interventions. Future research could incorporate additional factors such as global trade and sector-specific impacts to provide a more comprehensive view of the recovery dynamics. The findings highlight that while fiscal policies played a crucial role, the effectiveness of these interventions varied significantly, underscoring the need for tailored approaches based on the economic context of each country.